\documentstyle[12pt,epic,eepic]{article}

\makeatletter
        \def\@cite#1#2{\leavevmode\hbox{$^{\mbox{\the\scriptfont0 #1}}$}}
\makeatother

\font\tenrm=cmr10

\font\elevenbf=cmbx10 scaled\magstep 1
\font\elevenrm=cmr10 scaled\magstep 1
\font\elevenit=cmti10 scaled\magstep 1

     \def\sD{\scriptscriptstyle D}

     \def\half{ {1 \over 2} }
     \def\betamu{ {\beta\mu} }

\renewenvironment{thebibliography}[1]
 { \elevenrm
   \begin{list}{\arabic{enumi}.}
    {\usecounter{enumi} \setlength{\parsep}{0pt}
     \setlength{\itemsep}{3pt} \settowidth{\labelwidth}{#1.}
     \sloppy
    }}{\end{list}}

\begin{document}
\begin{flushright}
\begin{tabular}{l}
HUPD-9527 \hspace{0.5cm}\\
November 1995
\end{tabular}
\end{flushright}
\vglue 23pt
\begin{center}{{\bf PHASE STRUCTURE OF A FOUR-FERMION THEORY\\
               \vglue 10pt
               AT FINITE TEMPERATURE AND CHEMICAL POTENTIAL\footnote{
Talk presented at the Fourth International Workshop on Thermal Field
Theory and Their Application,  Dalian, P. R. China, 5--10 August, 1995.
The main part of this talk is based on the work in collaboration with
T.~Kouno and T.~Muta.\cite{IKM}
}\\}
\vglue 1.5cm
{Tomohiro Inagaki\\}
\baselineskip=30pt
{\it Department of Physics, Hiroshima University, \\}
\baselineskip=20pt
{\it Higashi-Hiroshima, Hiroshima 739, Japan \\}
\vglue 3cm
{\rm ABSTRACT}}
\end{center}
\vglue 0.6cm
{\rightskip=3pc
 \leftskip=3pc
 \rm\baselineskip=17pt
 \noindent
We discuss the chiral symmetry restoration at high temperature
and chemical potential for a four-fermion interaction theory
in arbitrary dimensions ($2\leq D<4$).
To investigate the ground state of the theory we calculate the effective
potential and the gap equation using the method of the $1/N$ expansion.
We study the phase structure at finite temperature and chemical potential
and show critical curves which divide the symmetric and
asymmetric phase.
We find that the first and second order phase transition coexist
for $2\leq D \leq 3$ depending on the value of the temperature and
chemical potential.
On the other hand only the second order phase transition is realized for
$3<D<4$.
We also present the critical behavior of the dynamically generated fermion
mass.
\vglue 0.6cm}
\newpage
{\elevenbf\noindent 1. Introduction}
\vglue 0.4cm
\baselineskip=22pt
After the pioneering work by Y.~Nambu and G.~Jona-Lasinio,\cite{NJL}
the idea of dynamical symmetry breaking caused by the non-vanishing
expectation value of the fermion and anti-fermion has played a decisive role
in modern particle physics.
The idea was introduced to discuss the chiral symmetry breaking in QCD
and investigate the phenomena of hadrons.
On the other hand the basic ingredient of the standard electroweak
theory is the spontaneous breaking of the gauge symmetry
$SU(2)\otimes U(1)$ and grand unified theories are constructed on
the basis of the Higgs mechanism.
We ordinarily consider the Higgs field is an elementary scalar field.
The dynamics of the Higgs fields, however, has not been well understood.
There is a possibility that the Higgs fields may be constructed
as bound states of fundamental fermions and the gauge symmetry is also
broken down dynamically.\cite{TC}

I am interested in studying the dynamical origin of the symmetry
breaking.
One of the possible environments where the models of dynamical symmetry
breaking may be tested is found in the early universe where the symmetry
of the  primary unified theory is broken down to yield lower level theories.
In the early universe it is not adequate to neglect the effect of the
curvature, temperature and chemical potential.
In this talk I discuss a simple toy model of the dynamical symmetry
breaking, Gross-Neveu type model,\cite{GN} at finite temperature and
chemical potential in arbitrary dimensions.
There are many pioneering works in this field.
In Ref. $5$ the critical temperature of the Gross-Neveu model was found
at the large $N$ limit.
In Ref. $6$ the Gross-Neveu model was discussed at finite chemical potential.
The Gross-Neveu type model has been studied at finite temperature and
chemical potential in two and three dimensions.${}^{7 \sim 9}$
Curvature effects also investigated in arbitrary dimensions.\cite{CURV}

I would like to report our investigation in a simple model
of the dynamical symmetry breaking at finite temperature and chemical
potential.
In Sec. $2$ I will briefly review the general properties of the
four-fermion theory, a simple model of the dynamical symmetry breaking,
for vanishing temperature and chemical potential.
In Sec. $3$ I will introduce the temperature and chemical potential in the
theory and investigate the phase structure with varying temperature and
chemical potential.
Section $4$ gives concluding remarks.
\vglue 0.6cm
{\elevenbf\noindent 2. Four-Fermion Theory}
\vglue 0.4cm
The four-fermion interaction theory is one of the prototype models
of the dynamical symmetry breaking.
In this talk we consider the simple four-fermion interaction theory with
$N$-components fermions described by the Lagrangian\cite{GNRENG}
\begin{equation}
     {\cal L} =
     \sum^{N}_{k=1}\bar{\psi}_{k}i\gamma_{\mu}\partial^{\mu}\psi_{k}
     +\frac{\lambda_0}{2N}\sum^{N}_{k=1}(\bar{\psi}_{k}\psi_{k})^{2}\, ,
\label{l:gn}
\end{equation}
where the index $k$ represents the flavors of the fermion field $\psi$
and $\lambda_{0}$ is a bare coupling constant.
In the following discussions, for simplicity, we neglect the flavor index.
In two space-time dimensions the theory is nothing but the Gross-Neveu
model.\cite{GN}
The theory has a discrete chiral symmetry,
$\bar\psi\psi \rightarrow -\bar\psi\psi$, and a global $SU(N)$ flavor
symmetry.
The discrete chiral symmetry prevents the Lagrangian to have mass terms.
Under the circumstance of the global $SU(N)$ symmetry we may work in the
scheme of the $1/N$ expansion.

In practical calculations it is more convenient to introduce a auxiliary
field $\sigma$ and start with the Lagrangian defined by,
\begin{equation}
     {\cal L} = \bar{\psi}i\gamma_{\mu}\partial^{\mu}\psi
     -\frac{N}{2\lambda_0}\sigma^{2}-\bar{\psi}\sigma\psi\, .
\label{l:yukawa}
\end{equation}
Replacing $\sigma$ in the Eq.$(\ref{l:yukawa})$ by the solution
of the equation of motion
$\displaystyle \sigma\sim -\lambda_{0}\bar{\psi}\psi/N$,
the Eq.$(\ref{l:gn})$ is reproduced.
If the non-vanishing vacuum expectation value is assigned to the auxiliary
field $\sigma$ then there appears a mass term for the fermion field $\psi$
and the discrete chiral symmetry is eventually broken.

We would like to find a ground state of the system described
by the four-fermion theory.
For this purpose we evaluate an effective potential at the large $N$ limit.
The grand state of the theory is determined by observing the minimum of
the effective potential.
In the leading order of the $1/N$ expansion the effective potential
is given by
\begin{equation}
\begin{array}{rcl}
     V_{0}(\sigma ) &=& \displaystyle\frac{1}{2\lambda_0}\sigma^{2}
                      +i\, \mbox{ln\ det}
                      (i \gamma_{\mu}\partial^{\mu}-\sigma)
                      +\mbox{O}(1/N)\\
                    &=& \displaystyle\frac{1}{2\lambda_0}\sigma^{2}
                      -\frac{1}{(2\pi)^{\sD/2}D}
                      \Gamma \left( 1-\frac{D}{2} \right)\sigma^{\sD} \, .
\label{v:gn}
\end{array}
\end{equation}
It should be noted that the effective potential is normalized so that
$V_{0}(0)=0$.

The effective potential given in Eq.$(\ref{v:gn})$ is divergent in two
and four space-time dimensions.
In the case of the two space-time dimensions the four-fermion theory is
renormalizable, so that we can avoid the problem of divergences by the
usual renormalization procedure.
We define the renormalized coupling constant $\lambda$ by using the
renormalization condition
\begin{equation}
     \left.
     \frac{\partial^{2}V_{0}(\sigma)}{\partial \sigma^{2}}
     \right|_{\sigma = \sigma_{0}}
     =\frac{\sigma_{0}^{\sD-2}}{\lambda}\, ,
\label{cond:ren}
\end{equation}
where $\sigma_{0}$ is the renormalization scale.
To satisfy the condition given in Eq.$(\ref{cond:ren})$ the renormalized
coupling $\lambda$ reads
\begin{equation}
     \frac{1}{\lambda_0}=\frac{1}{\lambda}\sigma_{0}^{\sD-2}
                       +\frac{1}{(2\pi)^{\sD/2}}(D-1)
                        \Gamma \left( 1-\frac{D}{2} \right)
                        \sigma_{0}^{\sD-2}\, .
\label{eqn:ren}
\end{equation}
Replacing the bare coupling constant with the renormalized one
the effective potential is no longer divergent in the
whole range of the space-time dimensions considered here: $2 \leq D < 4$ .

For $D=4$  the four-fermion theory is not renormalizable and hence we
can not cancel out the divergence by the renormalization procedure.
We regard the theory for $D=4-\epsilon$ with the sufficiently small and
positive $\epsilon$ as the regularization in four space-time dimensions
and consider the theory for $D=4$ as a low energy effective theory
stemming from more fundamental theories.

Evaluating the effective potential $V_{0}(\sigma)$ we find the phase
structure of the four-fermion theory.
If the coupling constant $\lambda$ is no less than a critical value
$\lambda_{cr}$ given by
\begin{equation}
     \lambda_{cr} = (2\pi)^{\sD/2}
                        \left[
                        (1-D)\Gamma \left( 1-\frac{D}{2} \right)
                        \right]^{-1}\, ,
\label{ccup:d}
\end{equation}
the shape of the effective potential is of a double well
and the minimum is located at non-vanishing $\sigma$.
In this case the discrete chiral symmetry of the theory is broken down
dynamically.
Evaluating the gap equation defined by
\begin{equation}
     \left.
     \frac{\partial V_{0}(\sigma)}{\partial \sigma}
     \right|_{\sigma = m}
     =0\, ,
\end{equation}
we find the dynamical mass of the fermion as a function of the coupling
constant $\lambda$,
\begin{equation}
     m = \sigma_0
   \left[
                \frac{(2\pi)^{\sD/2}}
           {\Gamma \left( 1-{\displaystyle{ D \over 2}} \right)}
                \left(
                \frac{1}{\lambda}-\frac{1}{\lambda_{cr}}
                \right)
          \right]^{1/(2-\sD)} \, .
\label{mass:d}
\end{equation}
As is well known, the shape of the effective potential $V_{0}(\sigma)$
is of a single well for $\lambda < \lambda_{cr}$. Thus the ground
state is invariant under the discrete chiral transformation.

In the following discussions we fix the coupling constant $\lambda$
above the critical coupling and see whether the chiral symmetry
is restored in an environment of the high temperature and chemical potential.
\vglue 0.6cm
{\elevenbf\noindent 3. Phase Structure at Finite Temperature and
Chemical Potential}
\vglue 0.4cm
Here I will introduce the temperature and chemical potential in the theory
and investigate the phase structure with varying the temperature and chemical
potential.
The $n$-points thermal Green function is defined by
\begin{equation}
     G^{\beta \mu}_{n}
     = \frac{\sum_{\alpha} e^{-\beta(E_{\alpha}-\mu N_{\alpha})}
             \langle \alpha | \mbox{T}(\psi(x_{1}),\,\cdots ,\,\psi(x_{i}),
             \bar{\psi}(x_{i+1}),\,\cdots ,\,\bar{\psi}(x_{n}))
             | \alpha \rangle}
            {\sum_{\alpha} e^{-\beta(E_{\alpha}-\mu N_{\alpha})}}
\,
,
\label{eqn:gfunc}
\end{equation}
where $E_\alpha$ and
\ $N_\alpha$ are the energy and particle number in the state
specified by a quantum number $\alpha$ respectively,
$\beta=1/kT$ with $k$ the Boltzmann
constant and $T$ the temperature
and $\mu$ is the chemical potential.
Using the path integral formalism the partition function for
the Green function $(\ref{eqn:gfunc})$ reads\cite{EFFT}
\begin{equation}
     Z^{\beta\mu}=\int [d\psi][d\bar{\psi}]\ \mbox{exp}\
                    \left[i\int^{-i\beta}_{0}dt\int d^{\sD-1} x
                    ({\cal L}+\mu N)\right]\, ,
\end{equation}
where $N$ is the number operator.

Following the standard procedure of the Matsubara Green function, we
calculate the effective potential of the theory in the leading order
of the $1/N$ expansion and find
\begin{equation}
     V(\sigma)=V_{0}(\sigma) + V^{\beta \mu}(\sigma)\, ,
\label{v:dev}
\end{equation}
where $V_{0}(\sigma)$ is the effective potential
for $T=\mu=0$ shown in Eq.$(\ref{v:gn})$ and
$V^{\beta \mu}(\sigma)$ is given by
\begin{eqnarray}
     V^{\beta \mu}(\sigma) & = &
     -\frac{1}{\beta}\frac{\sqrt{2}}{(2\pi)^{(\sD-1)/2}}
      \frac{1}{\Gamma \left(\displaystyle\frac{D-1}{2}\right)}
                                                  \nonumber \\
      & & \times\int dk k^{\sD-2}\left[ \mbox{ln}
      \frac{1+ e^{-\beta(\sqrt{k^{2}+\sigma^{2}}+\mu)}}
           {1+ e^{-\beta(k+\mu)}}
    + \ln
      \frac{1+ e^{-\beta(\sqrt{k^{2}+\sigma^{2}}-\mu)}}
           {1+ e^{-\beta(k-\mu)}} \right] \, .
\label{v:exp2}
\end{eqnarray}
As the influence from the finite temperature and chemical potential (i.e.
$V^{\beta\mu}(\sigma)$) is not divergent, we can use the same
renormalization procedure presented in the previous section to remove
the divergence.
 The renormalized effective potential is obtained by replacing the coupling
constant $\lambda_{0}$ with the renormalized one $\lambda$.

To study the phase structure at finite temperature and chemical
potential we evaluate the effective potential $(\ref{v:dev})$ with varying
the temperature and chemical potential.
The dynamical fermion mass is obtained by the vacuum expectation
value of the auxiliary field $\sigma$.
We can find it by observing the minimum of the effective potential.
The necessary condition for the minimum is given by the gap equation:
\begin{equation}
     \left.
     \frac{\partial V(\sigma)}{\partial \sigma}
     \right|_{\sigma = m_{\beta\mu}} = m_{\beta\mu}
     f(m_{\beta\mu}, \beta, \mu)=0\, .
\label{eq:gap}
\end{equation}
The dynamical fermion mass is obtained by the non-trivial solution of
the gap equation.
The dynamical fermion mass smoothly disappears at the critical point
for the second order phase transition.
The point is given by $f(0, \beta, \mu) = 0$.
For the first order phase transition the effective potential has the same
value at two local minimums.
Thus the critical point is obtained by the solution of the equations:
\begin{equation}
     f(m_{\beta\mu}, \beta, \mu) = 0\, ,\mbox{\hspace{2ex}}
     V(m_{\beta\mu}) = V(0) = 0\, .
\end{equation}

We perform the full analysis of this type for $2\leq D < 4$
and obtain the critical curves which divide the chiral
symmetric and asymmetric phase on $T-\mu$ plane.
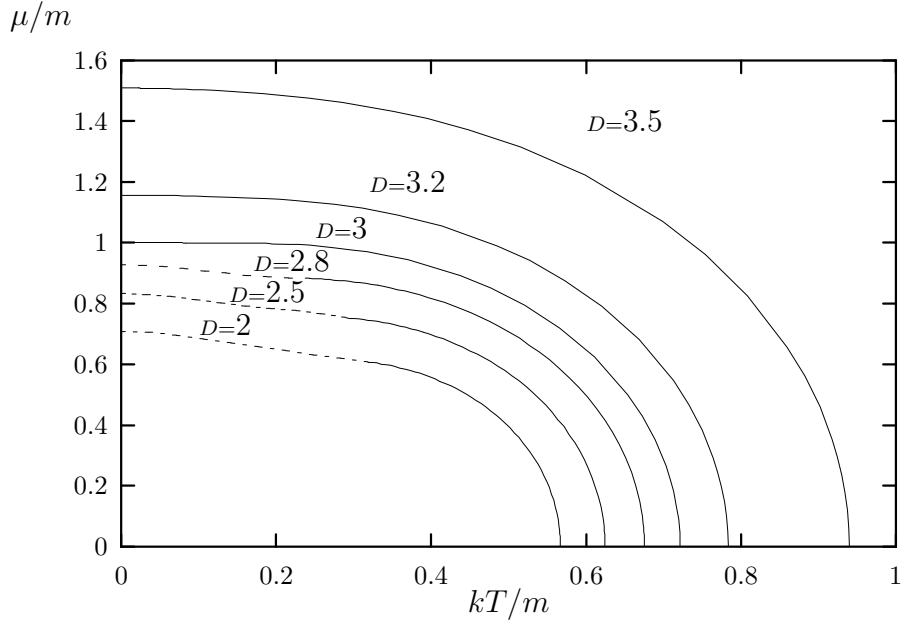
\begin{figure}
\setlength{\unitlength}{0.240900pt}
\begin{picture}(1500,900)(0,0)
\tenrm
\thicklines \path(220,113)(240,113)
\thicklines \path(1436,113)(1416,113)
\put(198,113){\makebox(0,0)[r]{0}}
\thicklines \path(220,209)(240,209)
\thicklines \path(1436,209)(1416,209)
\put(198,209){\makebox(0,0)[r]{0.2}}
\thicklines \path(220,304)(240,304)
\thicklines \path(1436,304)(1416,304)
\put(198,304){\makebox(0,0)[r]{0.4}}
\thicklines \path(220,400)(240,400)
\thicklines \path(1436,400)(1416,400)
\put(198,400){\makebox(0,0)[r]{0.6}}
\thicklines \path(220,495)(240,495)
\thicklines \path(1436,495)(1416,495)
\put(198,495){\makebox(0,0)[r]{0.8}}
\thicklines \path(220,591)(240,591)
\thicklines \path(1436,591)(1416,591)
\put(198,591){\makebox(0,0)[r]{1}}
\thicklines \path(220,686)(240,686)
\thicklines \path(1436,686)(1416,686)
\put(198,686){\makebox(0,0)[r]{1.2}}
\thicklines \path(220,782)(240,782)
\thicklines \path(1436,782)(1416,782)
\put(198,782){\makebox(0,0)[r]{1.4}}
\thicklines \path(220,877)(240,877)
\thicklines \path(1436,877)(1416,877)
\put(198,877){\makebox(0,0)[r]{1.6}}
\thicklines \path(220,113)(220,133)
\thicklines \path(220,877)(220,857)
\put(220,68){\makebox(0,0){0}}
\thicklines \path(463,113)(463,133)
\thicklines \path(463,877)(463,857)
\put(463,68){\makebox(0,0){0.2}}
\thicklines \path(706,113)(706,133)
\thicklines \path(706,877)(706,857)
\put(706,68){\makebox(0,0){0.4}}
\thicklines \path(950,113)(950,133)
\thicklines \path(950,877)(950,857)
\put(950,68){\makebox(0,0){0.6}}
\thicklines \path(1193,113)(1193,133)
\thicklines \path(1193,877)(1193,857)
\put(1193,68){\makebox(0,0){0.8}}
\thicklines \path(1436,113)(1436,133)
\thicklines \path(1436,877)(1436,857)
\put(1436,68){\makebox(0,0){1}}
\thicklines \path(220,113)(1436,113)(1436,877)(220,877)(220,113)
\put(45,945){\makebox(0,0)[l]{\shortstack{$\mu/m$}}}
\put(828,23){\makebox(0,0){$kT/m$}}
\put(342,462){\makebox(0,0)[l]{${\scriptstyle D=}2$}}
\put(389,514){\makebox(0,0)[l]{${\scriptstyle D=}2.5$}}
\put(427,562){\makebox(0,0)[l]{${\scriptstyle D=}2.8$}}
\put(524,614){\makebox(0,0)[l]{${\scriptstyle D=}3$}}
\put(609,686){\makebox(0,0)[l]{${\scriptstyle D=}3.2$}}
\put(950,782){\makebox(0,0)[l]{${\scriptstyle D=}3.5$}}
\thinlines \path(909,113)(909,113)(909,113)(909,114)(909,114)(909,114)
(909,115)(909,115)(909,116)(909,116)(909,117)(909,118)(909,120)(909,121)
(909,124)(909,126)(909,129)(908,134)(908,139)(907,145)(906,155)(904,165)
(901,176)(899,186)(895,196)(892,206)(888,215)(883,225)(879,234)(874,243)
(868,252)(862,261)(856,269)(850,277)(844,285)(837,292)(830,300)(823,307)
(815,314)(808,320)(800,326)(793,332)(785,338)(777,343)(769,348)(761,353)
(753,358)(745,362)(737,366)(728,370)(720,373)
\thinlines \path(720,373)(712,377)(704,380)(696,383)(688,386)(681,388)
(673,390)(665,392)(657,394)(650,396)(643,398)(635,399)(628,400)(621,402)
(614,403)(607,403)
\thinlines \dashline[3]{8}(220,451)(263,450)(306,445)(349,439)(392,433)
(435,427)(478,421)(521,414)(564,409)(607,403)
\thinlines \path(979,113)(979,113)(979,113)(979,114)(979,114)(979,115)
(979,115)(979,116)(979,117)(979,118)(979,119)(979,121)(979,123)(979,125)
(979,129)(979,133)(978,137)(978,145)(977,153)(976,160)(973,176)(969,191)
(964,206)(959,221)(953,236)(947,250)(940,263)(932,277)(923,289)(915,302)
(905,313)(896,325)(886,336)(876,346)(865,356)(855,365)(844,374)(833,382)
(822,390)(811,397)(800,404)(789,410)(778,416)(767,421)(756,427)(745,431)
(735,436)(725,440)(714,443)(704,447)(694,450)
\thinlines \path(694,450)(685,453)(675,455)(666,458)(657,460)(648,462)
(640,464)(631,465)(623,467)(615,468)(607,469)(600,470)(593,471)(585,472)
(578,472)(572,473)
\thinlines \dashline[3]{8}(220,511)(259,509)(298,506)(337,501)(376,496)
(415,491)(454,487)(494,484)(533,479)(572,473)
\thinlines \path(1041,113)(1041,113)(1041,114)(1041,114)(1041,115)
(1041,116)(1041,117)(1041,119)(1041,120)(1041,122)(1041,125)(1041,128)
(1041,131)(1040,137)(1040,143)(1039,149)(1038,161)(1036,172)(1034,184)
(1028,207)(1021,230)(1012,252)(1002,273)(991,293)(979,312)(966,330)
(952,348)(938,364)(923,379)(907,394)(892,407)(876,419)(860,430)(845,441)
(829,450)(813,459)(798,467)(783,474)(769,481)(754,487)(740,492)(727,497)
(713,501)(701,505)(688,509)(676,512)(665,515)(654,517)(643,520)(632,522)
(622,524)
\thinlines \path(622,524)(613,525)(603,527)(594,528)(586,529)(578,530)
(570,531)(562,531)(554,532)(547,533)(540,533)(534,534)(527,534)(521,534)
(515,535)
\thinlines \dashline[3]{8}(220,556)(253,555)(286,553)(318,550)(351,546)
(384,543)(417,539)(449,537)(482,535)(515,535)
\thinlines \path(1097,113)(1097,113)(1097,135)(1095,157)(1091,178)
(1087,200)(1081,221)(1075,241)(1058,281)(1038,318)(1015,352)(963,411)
(908,457)(853,493)(800,520)(752,539)(670,564)(636,571)(606,577)(579,580)
(555,583)(515,587)(499,588)(484,588)(458,589)(447,590)(436,590)(427,590)
(418,590)(403,590)(396,590)(390,590)(384,590)(379,590)(374,590)(369,590)
(364,590)(360,590)(356,590)(352,590)(349,590)(345,590)(342,590)(339,590)
(336,590)(333,590)(331,590)(328,590)(326,590)(324,590)(321,590)
\thinlines \path(321,590)(319,590)(317,590)(315,591)(313,591)(312,591)
(310,591)(308,591)(307,591)(305,591)(304,591)(302,591)(301,591)(299,591)
(298,591)(297,591)(296,591)(294,591)(293,591)(292,591)(291,591)(290,591)
(289,591)(288,591)(287,591)(286,591)(285,591)(284,591)(284,591)(283,591)
(282,591)(281,591)(280,591)(280,591)(279,591)(278,591)(277,591)(277,591)
(276,591)(275,591)(275,591)(274,591)(274,591)(273,591)(272,591)(272,591)
(271,591)(271,591)(270,591)(270,591)(269,591)
\thinlines \path(269,591)(269,591)(268,591)(268,591)(267,591)(267,591)
(266,591)(266,591)(265,591)(265,591)(265,591)(264,591)(264,591)(263,591)
(263,591)(263,591)(262,591)(262,591)(261,591)(261,591)(261,591)(260,591)
(260,591)(260,591)(259,591)(259,591)(259,591)(259,591)(259,591)(259,591)
(258,591)(258,591)(258,591)(258,591)(258,591)(258,591)(258,591)(258,591)
(258,591)(258,591)(258,591)(258,591)(258,591)(257,591)(257,591)(257,591)
(257,591)(257,591)(257,591)(257,591)(257,591)
\thinlines \path(257,591)(256,591)(256,591)(256,591)(256,591)(256,591)
(256,591)(256,591)(256,591)(256,591)(255,591)(253,591)(251,591)(250,591)
(249,591)(248,591)(248,591)(248,591)(248,591)(247,591)(247,591)(247,591)
(247,591)(247,591)(247,591)(247,591)(247,591)(247,591)(247,591)(247,591)
(246,591)(246,591)(246,591)(246,591)(246,591)(246,591)(245,591)(245,591)
(245,591)(245,591)(245,591)(245,591)(245,591)(245,591)(245,591)(245,591)
(245,591)(245,591)(245,591)(245,591)(245,591)
\thinlines \path(245,591)(244,591)(244,591)(244,591)(244,591)(244,591)
(244,591)(244,591)(244,591)(244,591)(244,591)(244,591)(243,591)(243,591)
(242,591)(241,591)(241,591)(240,591)(240,591)(240,591)(240,591)(240,591)
(240,591)(240,591)(240,591)(240,591)(240,591)(240,591)(240,591)(240,591)
(240,591)(240,591)(240,591)(240,591)(239,591)(239,591)(239,591)(239,591)
(239,591)(239,591)(239,591)(239,591)(239,591)(239,591)(239,591)(239,591)
(239,591)(239,591)(239,591)(238,591)(238,591)
\thinlines \path(238,591)(237,591)(237,591)(236,591)(236,591)(236,591)
(236,591)(235,591)(235,591)(235,591)(235,591)(235,591)(235,591)(235,591)
(235,591)(235,591)(235,591)(235,591)(235,591)(235,591)(235,591)(235,591)
(235,591)(235,591)(235,591)(235,591)(235,591)(235,591)(235,591)(235,591)
(235,591)(235,591)(235,591)(235,591)(235,591)(235,591)(235,591)(234,591)
(234,591)(234,591)(234,591)(234,591)(234,591)(234,591)(234,591)(234,591)
(234,591)(234,591)(234,591)(234,591)(234,591)
\thinlines \path(234,591)(234,591)(234,591)(234,591)(234,591)(234,591)
(234,591)(234,591)(234,591)(234,591)(234,591)(233,591)(233,591)(233,591)
(233,591)(233,591)(233,591)(233,591)(233,591)(233,591)(233,591)(233,591)
(233,591)(233,591)(233,591)(233,591)(233,591)(233,591)(232,591)(232,591)
\thinlines \path(1173,113)(1173,113)(1172,137)(1170,161)(1167,184)
(1162,207)(1156,230)(1149,253)(1132,296)(1111,336)(1087,374)(1033,439)
(974,491)(916,531)(861,563)(810,586)(723,617)(654,634)(599,645)(555,651)
(492,657)(468,659)(447,660)(416,661)(391,662)(357,663)(335,664)(318,664)
(306,665)(296,665)(289,665)(283,665)(277,665)(273,665)(269,665)(266,665)
(263,665)(261,665)(258,665)(256,665)(254,665)(253,665)(251,665)(250,665)
(249,665)(248,665)(246,665)(246,665)(245,665)(244,665)(243,665)
\thinlines \path(243,665)(242,665)(242,665)(241,665)(240,665)(240,665)
(239,665)(239,665)(238,665)(238,665)(237,665)(237,665)(236,665)(236,665)
(236,665)(235,665)(235,665)(235,665)(234,665)(234,665)(220,665)
\thinlines \path(1363,113)(1363,113)(1362,142)(1360,170)(1356,198)
(1351,226)(1344,254)(1336,281)(1317,333)(1293,382)(1265,427)(1203,507)
(1137,572)(1070,624)(948,697)(847,741)(765,768)(700,786)(647,797)(568,811)
(513,818)(442,825)(398,828)(369,830)(348,831)(332,831)(320,832)(310,832)
(302,832)(295,833)(289,833)(284,833)(280,833)(276,833)(273,833)(270,833)
(267,833)(265,833)(263,833)(261,833)(259,833)(257,833)(256,833)(255,833)
(253,833)(252,833)(251,833)(250,833)(249,834)(248,834)(247,834)
\thinlines \path(247,834)(246,834)(246,834)(245,834)(244,834)(244,834)
(243,834)(242,834)(242,834)(241,834)(241,834)(240,834)(240,834)(240,834)
(239,834)(239,834)(238,834)(220,834)
\end{picture}
\caption{The critical curves for $2\leq D < 4$ .
The dashed lines represent the first order phase transition while
the full lines represent the second order phase transition.
}
\label{fig:grad}
\end{figure}
In Fig. $1$ the critical curves on $T-\mu$ plane are presented.
In drawing Fig. $1$ the temperature and the chemical potential
are normalized by the dynamical fermion mass $m$ at $T=\mu=0$.
Thus we find that the chiral symmetry is restored in an environment
of the high temperature and chemical potential for $2 \leq D < 4$.
{}From Fig. $1$ we clearly see that only the second order phase
transition is realized for the space-time dimensions $3 < D < 4$
while the first order phase transition also exist for $2\leq D \leq 3$.
For $D=2$ Fig. $1$ reproduces that obtained by U.~Wolff.\cite{D2GN}
In three space-time dimensions Fig. $1$ agrees with that obtained
by K.~Klimenko.\cite{D3GN}

Next we show the critical behavior of the dynamical fermion mass.
The dynamical fermion mass is obtained by the vacuum expectation
value of the auxiliary field $\sigma$.
We calculate the minimum of the effective potential numerically.
In Fig.~$2$ we plot the dynamical fermion mass $m_{\beta\mu}$ as the
function of chemical potential $\mu$ with temperature $kT/m$ fixed at
$0, 0.2$ and $0.4$.
It is obviously observed in Fig. $2$, the mass gap appears at low
temperature for $D=2.5$.
In the space-time dimensions greater than three there is no mass gap.
The mass gap exists only for $T=0$ in three space-time dimensions.
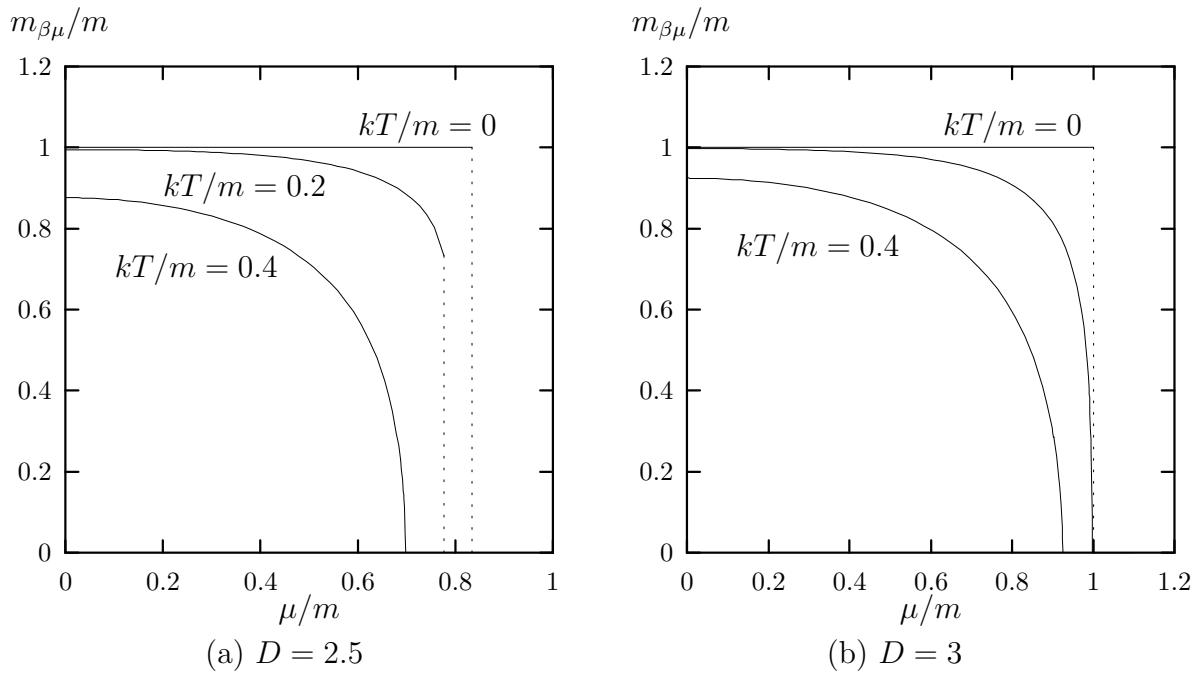
\begin{figure}
\hspace*{-3em}
   \begin{minipage}[t]{.47\linewidth}
\setlength{\unitlength}{0.240900pt}
\begin{picture}(1049,900)(0,0)
\tenrm
\thicklines \path(220,113)(240,113)
\thicklines \path(985,113)(965,113)
\put(198,113){\makebox(0,0)[r]{0}}
\thicklines \path(220,240)(240,240)
\thicklines \path(985,240)(965,240)
\put(198,240){\makebox(0,0)[r]{0.2}}
\thicklines \path(220,368)(240,368)
\thicklines \path(985,368)(965,368)
\put(198,368){\makebox(0,0)[r]{0.4}}
\thicklines \path(220,495)(240,495)
\thicklines \path(985,495)(965,495)
\put(198,495){\makebox(0,0)[r]{0.6}}
\thicklines \path(220,622)(240,622)
\thicklines \path(985,622)(965,622)
\put(198,622){\makebox(0,0)[r]{0.8}}
\thicklines \path(220,750)(240,750)
\thicklines \path(985,750)(965,750)
\put(198,750){\makebox(0,0)[r]{1}}
\thicklines \path(220,877)(240,877)
\thicklines \path(985,877)(965,877)
\put(198,877){\makebox(0,0)[r]{1.2}}
\thicklines \path(220,113)(220,133)
\thicklines \path(220,877)(220,857)
\put(220,68){\makebox(0,0){0}}
\thicklines \path(373,113)(373,133)
\thicklines \path(373,877)(373,857)
\put(373,68){\makebox(0,0){0.2}}
\thicklines \path(526,113)(526,133)
\thicklines \path(526,877)(526,857)
\put(526,68){\makebox(0,0){0.4}}
\thicklines \path(679,113)(679,133)
\thicklines \path(679,877)(679,857)
\put(679,68){\makebox(0,0){0.6}}
\thicklines \path(832,113)(832,133)
\thicklines \path(832,877)(832,857)
\put(832,68){\makebox(0,0){0.8}}
\thicklines \path(985,113)(985,133)
\thicklines \path(985,877)(985,857)
\put(985,68){\makebox(0,0){1}}
\thicklines \path(220,113)(985,113)(985,877)(220,877)(220,113)
\put(133,945){\makebox(0,0)[l]{\shortstack{$m_{\beta\mu}/m$}}}
\put(602,23){\makebox(0,0){$\mu/m$}}
\put(679,782){\makebox(0,0)[l]{$kT/m=0$}}
\put(373,686){\makebox(0,0)[l]{$kT/m=0.2$}}
\put(297,559){\makebox(0,0)[l]{$kT/m=0.4$}}
\thinlines \path(220,750)(220,750)(247,750)(273,750)(300,750)(326,750)
(353,750)(380,750)(406,750)(433,750)(459,750)(486,750)(512,750)(539,750)
(566,750)(592,750)(619,750)(645,750)(672,750)(699,750)(725,750)(752,750)
(778,750)(805,750)(832,750)(858,750)
\thinlines \dashline[10]{5}(814,580)(814,113)
\thinlines \dashline[10]{5}(858,750)(858,113)
\thinlines \path(220,746)(220,746)(228,746)(231,746)(233,746)(239,746)
(245,746)(251,746)(263,746)(274,746)(286,746)(298,746)(309,746)(321,746)
(324,746)(325,746)(327,746)(327,746)(328,746)(328,746)(329,746)(329,746)
(330,746)(333,746)(344,745)(356,745)(368,745)(379,745)(391,744)(403,744)
(414,744)(426,743)(437,743)(449,742)(461,742)(472,741)(484,740)(496,740)
(507,739)(519,738)(531,737)(542,736)(554,735)(566,734)(577,732)(589,731)
(601,729)(612,727)(624,725)(636,723)(647,720)
\thinlines \path(647,720)(659,718)(671,715)(682,711)(694,707)(706,703)
(717,698)(729,692)(741,686)(752,678)(764,669)(776,658)(787,644)(799,624)
(814,580)
\thinlines \path(220,671)(220,671)(228,671)(231,671)(234,671)(240,671)
(246,670)(252,670)(265,670)(277,669)(289,668)(302,668)(314,666)(326,665)
(339,664)(351,662)(363,660)(376,658)(388,656)(400,654)(413,651)(425,648)
(437,645)(450,642)(462,638)(474,634)(487,630)(499,626)(511,621)(524,615)
(536,609)(548,603)(561,596)(573,589)(585,580)(598,571)(610,561)(622,550)
(635,538)(647,525)(659,509)(672,492)(684,472)(696,448)(709,419)(721,383)
(727,361)(733,335)(739,301)(743,282)(746,258)
\thinlines \path(746,258)(747,244)(749,228)(750,209)(751,204)(751,199)
(752,184)(753,141)(754,113)
\end{picture}
       \hspace*{9em}\mbox{(a) $D=2.5$}
   \end{minipage}
\hfill
   \begin{minipage}[t]{.53\linewidth}
\setlength{\unitlength}{0.240900pt}
\begin{picture}(1049,900)(0,0)
\tenrm
\thicklines \path(220,113)(240,113)
\thicklines \path(985,113)(965,113)
\put(198,113){\makebox(0,0)[r]{0}}
\thicklines \path(220,240)(240,240)
\thicklines \path(985,240)(965,240)
\put(198,240){\makebox(0,0)[r]{0.2}}
\thicklines \path(220,368)(240,368)
\thicklines \path(985,368)(965,368)
\put(198,368){\makebox(0,0)[r]{0.4}}
\thicklines \path(220,495)(240,495)
\thicklines \path(985,495)(965,495)
\put(198,495){\makebox(0,0)[r]{0.6}}
\thicklines \path(220,622)(240,622)
\thicklines \path(985,622)(965,622)
\put(198,622){\makebox(0,0)[r]{0.8}}
\thicklines \path(220,750)(240,750)
\thicklines \path(985,750)(965,750)
\put(198,750){\makebox(0,0)[r]{1}}
\thicklines \path(220,877)(240,877)
\thicklines \path(985,877)(965,877)
\put(198,877){\makebox(0,0)[r]{1.2}}
\thicklines \path(220,113)(220,133)
\thicklines \path(220,877)(220,857)
\put(220,68){\makebox(0,0){0}}
\thicklines \path(348,113)(348,133)
\thicklines \path(348,877)(348,857)
\put(348,68){\makebox(0,0){0.2}}
\thicklines \path(475,113)(475,133)
\thicklines \path(475,877)(475,857)
\put(475,68){\makebox(0,0){0.4}}
\thicklines \path(603,113)(603,133)
\thicklines \path(603,877)(603,857)
\put(603,68){\makebox(0,0){0.6}}
\thicklines \path(730,113)(730,133)
\thicklines \path(730,877)(730,857)
\put(730,68){\makebox(0,0){0.8}}
\thicklines \path(858,113)(858,133)
\thicklines \path(858,877)(858,857)
\put(858,68){\makebox(0,0){1}}
\thicklines \path(985,113)(985,133)
\thicklines \path(985,877)(985,857)
\put(985,68){\makebox(0,0){1.2}}
\thicklines \path(220,113)(985,113)(985,877)(220,877)(220,113)
\put(133,945){\makebox(0,0)[l]{\shortstack{$m_{\beta\mu}/m$}}}
\put(602,23){\makebox(0,0){$\mu/m$}}
\put(622,782){\makebox(0,0)[l]{$kT/m=0$}}
\put(297,591){\makebox(0,0)[l]{$kT/m=0.4$}}
\thinlines \path(221,748)(221,748)(221,748)(222,748)(222,748)(222,748)
(223,748)(224,748)(226,748)(228,748)(231,748)(235,748)(242,748)(249,748)
(256,748)(264,748)(278,748)(292,748)(306,748)(321,747)(335,747)(349,747)
(364,747)(378,746)(392,746)(407,746)(421,745)(435,745)(449,744)(464,744)
(478,743)(492,742)(507,741)(521,740)(535,739)(549,738)(564,736)(578,735)
(592,733)(607,730)(621,728)(635,725)(650,722)(664,718)(678,714)(692,709)
(707,703)(721,696)(735,688)(750,678)(764,666)
\thinlines \path(764,666)(778,652)(793,633)(800,621)(807,608)(814,592)
(821,572)(828,547)(832,532)(835,515)(839,494)(843,468)(846,435)(848,414)
(850,389)(851,372)(851,362)(852,357)(852,332)(853,305)(854,292)(854,272)
(855,220)(856,172)(856,113)(856,113)(857,113)(859,113)(860,113)(864,113)
(878,113)(893,113)(907,113)(921,113)
\thinlines \path(221,702)(221,702)(221,702)(222,702)(222,702)(222,702)
(223,702)(224,701)(226,701)(228,701)(231,701)(235,701)(242,701)(249,701)
(256,701)(264,701)(278,700)(292,699)(306,699)(321,697)(335,696)(349,695)
(364,693)(378,691)(392,689)(407,687)(421,684)(435,681)(449,678)(464,675)
(478,671)(492,667)(507,663)(521,658)(535,653)(549,647)(564,641)(578,634)
(592,626)(607,618)(621,609)(635,599)(650,588)(664,575)(678,561)(692,546)
(707,528)(721,507)(735,483)(750,454)(764,419)
\thinlines \path(764,419)(778,374)(785,346)(793,310)(794,300)(795,295)
(796,295)(796,291)(797,285)(798,278)(800,265)(803,234)(805,215)(807,188)
(809,153)(809,135)(810,113)
\thinlines \path(220,750)(220,750)(858,750)
\thinlines \dashline[10]{5}(858,750)(858,113)
\end{picture}
       \hspace*{9em}(b) $D=3$
   \end{minipage}
\caption{Dynamical fermion mass $m_\betamu$ as
           a function of the chemical
           potential $\mu$ with temperature $kT/m$ fixed
           at 0, 0.2, 0.4.}
\label{fig:chemmass}
\end{figure}

In Fig. $3$ we plot the dynamical fermion mass $m_{\beta\mu}$ as the
function of temperature $T$ with chemical potential $\mu /m$ fixed
at $0, 0.4$ and $0.8$.
For $D=2.5$ the mass gap appears at large chemical potential.
No mass gap is observed above $D=3$.

We are able to find analytically some specific points on the critical
curve.\cite{IKM}
\begin{figure}
\hspace*{-3em}
   \begin{minipage}[t]{.47\linewidth}
\setlength{\unitlength}{0.240900pt}
\begin{picture}(1049,900)(0,0)
\tenrm
\thicklines \path(220,113)(240,113)
\thicklines \path(985,113)(965,113)
\put(198,113){\makebox(0,0)[r]{0}}
\thicklines \path(220,240)(240,240)
\thicklines \path(985,240)(965,240)
\put(198,240){\makebox(0,0)[r]{0.2}}
\thicklines \path(220,368)(240,368)
\thicklines \path(985,368)(965,368)
\put(198,368){\makebox(0,0)[r]{0.4}}
\thicklines \path(220,495)(240,495)
\thicklines \path(985,495)(965,495)
\put(198,495){\makebox(0,0)[r]{0.6}}
\thicklines \path(220,622)(240,622)
\thicklines \path(985,622)(965,622)
\put(198,622){\makebox(0,0)[r]{0.8}}
\thicklines \path(220,750)(240,750)
\thicklines \path(985,750)(965,750)
\put(198,750){\makebox(0,0)[r]{1}}
\thicklines \path(220,877)(240,877)
\thicklines \path(985,877)(965,877)
\put(198,877){\makebox(0,0)[r]{1.2}}
\thicklines \path(220,113)(220,133)
\thicklines \path(220,877)(220,857)
\put(220,68){\makebox(0,0){0}}
\thicklines \path(329,113)(329,133)
\thicklines \path(329,877)(329,857)
\put(329,68){\makebox(0,0){0.1}}
\thicklines \path(439,113)(439,133)
\thicklines \path(439,877)(439,857)
\put(439,68){\makebox(0,0){0.2}}
\thicklines \path(548,113)(548,133)
\thicklines \path(548,877)(548,857)
\put(548,68){\makebox(0,0){0.3}}
\thicklines \path(657,113)(657,133)
\thicklines \path(657,877)(657,857)
\put(657,68){\makebox(0,0){0.4}}
\thicklines \path(766,113)(766,133)
\thicklines \path(766,877)(766,857)
\put(766,68){\makebox(0,0){0.5}}
\thicklines \path(876,113)(876,133)
\thicklines \path(876,877)(876,857)
\put(876,68){\makebox(0,0){0.6}}
\thicklines \path(985,113)(985,133)
\thicklines \path(985,877)(985,857)
\put(985,68){\makebox(0,0){0.7}}
\thicklines \path(220,113)(985,113)(985,877)(220,877)(220,113)
\put(133,945){\makebox(0,0)[l]{\shortstack{$m_{\beta\mu}/m$}}}
\put(602,23){\makebox(0,0){$kT/m$}}
\put(712,686){\makebox(0,0)[l]{$\mu/m =0$}}
\put(526,431){\makebox(0,0)[l]{$\mu/m =0.4$}}
\put(395,571){\makebox(0,0)[l]{$\mu/m =0.8$}}
\thinlines \path(220,750)(220,750)(234,749)(248,749)(262,749)(276,749)
(290,749)(304,749)(318,749)(388,749)(402,749)(416,748)(430,747)(444,746)
(458,745)(472,743)(486,741)(500,738)(514,735)(528,731)(542,727)(556,723)
(570,718)(584,712)(598,705)(612,698)(626,691)(640,682)(654,673)(668,663)
(682,652)(696,640)(710,627)(724,613)(738,598)(752,581)(766,563)
\thinlines \path(766,563)(766,563)(774,553)(781,543)(788,532)(795,520)
(802,508)(809,496)(817,483)(824,468)(831,453)(838,437)(845,420)(852,402)
(859,383)(867,361)(874,335)(881,303)(888,273)(895,232)(902,113)
\thinlines \path(220,750)(220,750)(326,749)(341,749)(356,748)(371,747)
(386,746)(401,744)(416,742)(431,739)(446,735)(461,731)(476,727)(491,722)
(506,716)(521,709)(537,702)(552,694)(567,686)(582,677)(597,666)(612,655)
(627,643)(642,629)(657,614)
\thinlines \path(657,614)(657,614)(664,608)(670,601)(676,594)(683,586)
(689,579)(696,571)(702,562)(708,554)(715,545)(721,536)(728,526)(734,517)
(740,506)(747,495)(753,484)(760,472)(766,459)(772,445)(779,431)(785,416)
(792,400)(798,382)(804,364)(811,344)(817,320)(824,291)(830,258)(836,219)
(843,113)
\thinlines \path(220,750)(220,750)(242,750)(264,749)(286,746)(307,738)
(329,726)(340,717)(351,706)(357,700)(358,698)(360,695)(361,694)(362,693)
(363,691)(364,691)(364,691)
\thinlines \dashline[10]{5}(364,691)(364,113)
\end{picture}
       \hspace*{9em}\mbox{(a) $D=2.5$}
   \end{minipage}
\hfill
   \begin{minipage}[t]{.53\linewidth}
\setlength{\unitlength}{0.240900pt}
\begin{picture}(1049,900)(0,0)
\tenrm
\thicklines \path(220,113)(240,113)
\thicklines \path(985,113)(965,113)
\put(198,113){\makebox(0,0)[r]{0}}
\thicklines \path(220,240)(240,240)
\thicklines \path(985,240)(965,240)
\put(198,240){\makebox(0,0)[r]{0.2}}
\thicklines \path(220,368)(240,368)
\thicklines \path(985,368)(965,368)
\put(198,368){\makebox(0,0)[r]{0.4}}
\thicklines \path(220,495)(240,495)
\thicklines \path(985,495)(965,495)
\put(198,495){\makebox(0,0)[r]{0.6}}
\thicklines \path(220,622)(240,622)
\thicklines \path(985,622)(965,622)
\put(198,622){\makebox(0,0)[r]{0.8}}
\thicklines \path(220,750)(240,750)
\thicklines \path(985,750)(965,750)
\put(198,750){\makebox(0,0)[r]{1}}
\thicklines \path(220,877)(240,877)
\thicklines \path(985,877)(965,877)
\put(198,877){\makebox(0,0)[r]{1.2}}
\thicklines \path(220,113)(220,133)
\thicklines \path(220,877)(220,857)
\put(220,68){\makebox(0,0){0}}
\thicklines \path(316,113)(316,133)
\thicklines \path(316,877)(316,857)
\put(316,68){\makebox(0,0){0.1}}
\thicklines \path(411,113)(411,133)
\thicklines \path(411,877)(411,857)
\put(411,68){\makebox(0,0){0.2}}
\thicklines \path(507,113)(507,133)
\thicklines \path(507,877)(507,857)
\put(507,68){\makebox(0,0){0.3}}
\thicklines \path(603,113)(603,133)
\thicklines \path(603,877)(603,857)
\put(603,68){\makebox(0,0){0.4}}
\thicklines \path(698,113)(698,133)
\thicklines \path(698,877)(698,857)
\put(698,68){\makebox(0,0){0.5}}
\thicklines \path(794,113)(794,133)
\thicklines \path(794,877)(794,857)
\put(794,68){\makebox(0,0){0.6}}
\thicklines \path(889,113)(889,133)
\thicklines \path(889,877)(889,857)
\put(889,68){\makebox(0,0){0.7}}
\thicklines \path(985,113)(985,133)
\thicklines \path(985,877)(985,857)
\put(985,68){\makebox(0,0){0.8}}
\thicklines \path(220,113)(985,113)(985,877)(220,877)(220,113)
\put(133,945){\makebox(0,0)[l]{\shortstack{$m_{\beta\mu}/m$}}}
\put(602,23){\makebox(0,0){$kT/m$}}
\put(736,686){\makebox(0,0)[l]{$\mu/m=0$}}
\put(335,444){\makebox(0,0)[l]{$\mu/m=0.8$}}
\thinlines \path(220,750)(220,750)(233,749)(246,749)(260,749)(273,749)
(286,749)(299,749)(312,750)(365,750)(378,749)(391,749)(405,748)(418,748)
(431,747)(444,745)(457,744)(471,742)(484,740)(497,737)(510,734)(523,731)
(537,727)(550,723)(563,718)(576,713)(589,707)(603,701)
\thinlines \path(603,701)(603,701)(610,697)(618,693)(626,689)(634,685)
(642,680)(650,675)(658,670)(666,664)(673,658)(681,652)(689,646)(697,639)
(705,631)(713,623)(721,615)(729,607)(736,598)(744,588)(752,579)(760,569)
(768,559)(776,548)(784,538)(792,527)(799,515)(807,504)(815,491)(823,477)
(831,462)(839,446)(847,429)(855,410)(863,389)(870,366)(878,339)(886,309)
(894,277)(902,222)(910,113)
\thinlines \path(220,750)(220,750)(233,750)(246,750)(312,750)(326,749)
(339,749)(352,748)(365,748)(378,747)(391,745)(405,744)(418,742)(431,740)
(444,737)(457,734)(471,731)(484,727)(497,723)(510,719)(523,713)(537,708)
(550,701)(563,695)(576,688)(589,680)(603,672)
\thinlines \path(603,672)(603,672)(609,668)(616,663)(623,659)(630,654)
(637,649)(644,643)(650,637)(657,631)(664,624)(671,617)(678,610)(685,602)
(692,595)(698,587)(705,580)(712,573)(719,565)(726,558)(733,549)(740,541)
(746,531)(753,521)(760,510)(767,499)(774,488)(781,476)(788,463)(794,449)
(801,435)(808,420)(815,403)(822,386)(829,367)(835,345)(842,320)(849,293)
(856,265)(863,213)(870,113)
\thinlines \path(220,750)(220,750)(266,750)(278,749)(289,747)(301,745)
(312,741)(324,738)(335,733)(347,728)(358,723)(370,717)(382,710)(393,703)
(405,696)(416,688)(428,679)(439,671)(451,661)(462,652)(474,642)(485,631)
(497,620)(509,609)(520,597)(532,585)(543,572)(555,557)
\thinlines \path(555,557)(555,557)(559,552)(563,547)(567,541)(571,535)
(575,529)(579,523)(583,517)(588,511)(592,504)(596,498)(600,491)(604,485)
(608,478)(612,472)(616,465)(620,458)(625,451)(629,444)(633,437)(637,430)
(641,422)(645,414)(649,406)(653,398)(657,389)(661,380)(666,370)(670,360)
(674,349)(678,338)(682,325)(686,312)(690,297)(694,281)(698,264)(703,247)
(707,222)(711,180)(715,113)
\end{picture}
       \hspace*{9em}(b) $D=3$
   \end{minipage}
\caption{Dynamical fermion
           mass $m_\betamu$ as a function
           of the temperature $T$ with the chemical
           potential $\mu/m$ fixed at 0, 0.4, 0.8.}
\label{fig:tmass}
\end{figure}
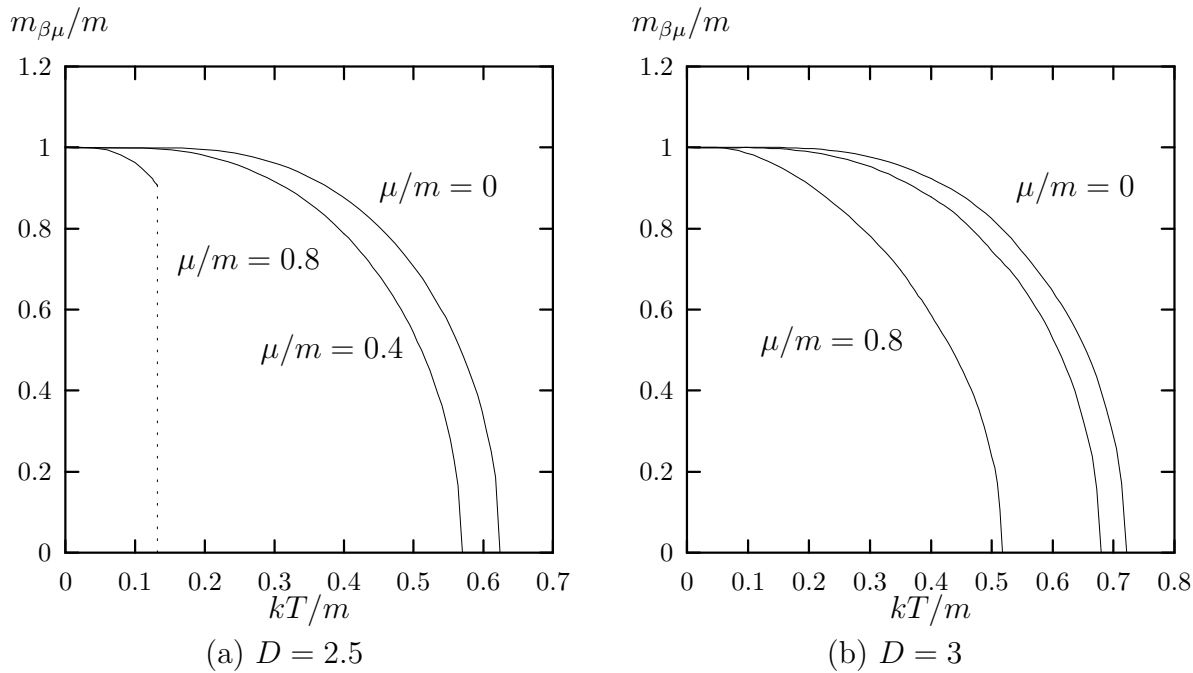
The critical temperature at $\mu =0$ is given by
\begin{equation}
   \beta_{cr} = \frac{2\pi}{m}
       \left[
           \frac{2\Gamma\left({\displaystyle{3-D \over 2}}\right)}
           {\sqrt{\pi}\Gamma
               \left({\displaystyle{2-D \over 2}}\right)}
           (2^{3-\sD}-1)\zeta(3-D)
       \right]^{1/(\sD-2)} \, ,
\label{cond:A}
\end{equation}
where $\zeta(z)$ is the Riemann zeta function.
Taking the two dimensional limit, the expression of the critical temperature
$(\ref{cond:A})$ reduces to the well-known formula shown in Ref. $5$.
For $D=3$ Eq.$(\ref{cond:A})$ reduces to the known formula in Ref. $9$.
For $2 \leq D \leq 3$ the critical chemical potential at $T=0$
reads
\begin{equation}
   \mu_{cr} = m\left[ {3 \over 4}
       {\rm B}\left({4-D \over 2},\,{D+1 \over 2}\right)
   \right]^{1/\sD} \, .
\label{cond:B}
\end{equation}
Taking the two dimensional limit, Eq.$(\ref{cond:B})$ reproduces the result
obtained in Ref. $8$.
For $3 < D < 4$ the critical chemical potential at $T=0$ is described by
\begin{equation}
   \mu_{cr} = m \left[
       \half{\rm B\,}\left({4-D \over 2},\,{D-1 \over 2}\right)
   \right]^{1/(\sD-2)} \, .
\label{cond:D}
\end{equation}
For $D=3$ Eq.$(\ref{cond:D})$ is agree with the known result.\cite{D3GN}
The critical temperature and the chemical potential at the boundary between
the first order and the second order phase transition is satisfied
\begin{equation}
   {\rm Re\,}\zeta\left(
       5-D,\,\half + i {\betamu \over 2\pi}
   \right)=0 \, ,
\label{cond:C}
\end{equation}
where $\zeta(z_{1},z_{2})$ is the generalized zeta function.
We obtain the  the boundary between the first order and the second order
phase transition by solving the Eq.$(\ref{cond:C})$.
The boundary is smoothly disappears at $D=3$ as is shown in Fig. $1$.

{\elevenbf\noindent 4. Conclusion}
\vglue 0.4cm
To study the thermodynamics of the dynamical symmetry breaking
is one of the crucial problems to test composite Higgs models.
In this talk we have considered the Gross-Neveu type model
as one of the prototype models of the dynamical symmetry breaking and
investigated the phase structure at finite temperature and chemical
potential in arbitrary dimensions.

Evaluating the effective potential in the leading order of the $1/N$
expansion, we found that the chiral symmetry restored for the sufficiently
high temperature and/or chemical potential in arbitrary
dimensions $2 \leq D < 4$.
We succeeded in finding the critical curves on $T-\mu$ plane
through analytical and numerical calculations of the effective potential.
Both the first order and the second order phase transitions coexist
for $2\leq D \leq 3$.
The first order phase transition was not realized for $D~>~3$.
The dynamical fermion mass was presented as a function of
the temperature $T$ or the chemical potential $\mu$.
At $D=2$ and $D=3$ the results derived in the present investigation
mostly reproduce the known results in the preceding works.
We obtained the analytical expressions of some critical points on the
critical curves.

For $D=2$ it is expected that the chiral symmetry is restored at any finite
value of the temperature in the case of finite $N$ through the creation
of a kink-antikink condensation.\cite{KINK}
In our method we can not deal with the influence of a space dependent
field configuration.
Kinks are, however, suppressed at the large $N$ limit and the phase
transition takes place at finite temperature.

We are interested in applying our result to phenomena in the early
universe and leave it future researches.
\vglue 0.6cm
{\elevenbf\noindent 5. Acknowledgements}
\vglue 0.4cm
The author would like to thank the organizers for their
hospitality during the conference.
Discussions with Taizo Muta and Toshihiro Kouno are gratefully
acknowledged.
I am indebted to the members of our Laboratory for
encouragements and discussions.
\vglue 0.6cm
{\elevenbf\noindent 6. References \hfil}
\vglue 0.4cm

\end{document}